\documentclass[namedreferences]{solarphysics}

\usepackage[hyperref,optionalrh]{spr-sola-addons} 
\usepackage{graphicx}        
\usepackage{color}           
\usepackage{breakurl}        

\usepackage{lmodern}

\usepackage{anyfontsize}

\usepackage{silence}

\usepackage[cp1251]{inputenc}

\chardef\us=`\_

\begin{document}

\begin{article}
\begin{opening}

\title{Occurrence rate of radio-loud and halo CMEs in solar cycle 25: Prediction using their correlation with sunspot numbers}

\author[addressref={aff1},email={ashanmugaraju@gmail.com}]{\fnm{A}~\lnm{Shanmugaraju}\orcid{0000-0002-2243-960X}}
\author[addressref=aff2,corref,email={kalai2230@gmail.com}]{\fnm{P}~\lnm{Pappa Kalaivani}\orcid{0000-0002-7364-0768}}
\author[addressref=aff3,email={moonyj@khu.ac.kr}]{\fnm{Y.-J}~\lnm{Moon}\orcid{0000-0001-6216-6944}}
\author[addressref={aff4,aff5}, email={prakash18941@gmail.com}]{\fnm{O}~\lnm{Prakash}\orcid{0000-0002-3312-0629}}
\address[id=aff1]{Department of Physics, Arul Anandar College, Karumathur, Madurai -625514, India}
\address[id=aff2]{Department of Physics, Ultra College of Engineering \& Technology for Women, Ultra Nagar, Madurai - 625 104, Tamil Nadu, India}
\address[id=aff3]{School of Space Research, Kyung Hee University, Yongin 446 -- 701, Republic of Korea}
\address[id=aff4]{Department of Physics, Sethu Institute of Technology, Pulloor, Kariapatti,Viruthunagar, Tamil Nadu -- 626 115, India}
\address[id=aff5]{Key Laboratory of Dark Matter and Space Astronomy, Purple Mountain Observatory, Chinese Academy of Sciences, Nanjing -- 210008, Jiangsu, China}

\runningauthor{Shanmugaraju et al.}
\runningtitle{\textit{Occurrence rate of radio-loud and halo CMEs in solar cycle 25}}

\begin{abstract}
The coronal mass ejections (CMEs) from the Sun are known for their space weather and geomagnetic consequences. Among all CMEs, so--called radio--loud (RL) and halo CMEs are considered the most energetic in the sense that they are usually faster and wider than the general population of CMEs.  Hence the study of RL and halo CMEs has become important and the prediction of their occurrence rate in a future cycle will give a warning in advance.  In the present paper, the occurrence rates of RL and halo CMEs in solar cycle (SC) 25 are predicted. For this, we obtained good correlations between the numbers of RL and halo CMEs in each year and the yearly mean sunspot numbers in the previous two cycles. The predicted values of sunspot numbers in SC 25 by NOAA/NASA were considered as representative indices and the corresponding numbers of RL and halo CMEs have been determined using linear relations.  Our results show that the maximum number of RL and halo CMEs will be around 39 $\pm$ 3 and 45 $\pm$ 4, respectively. Removing backside events, a set of front-side events was also considered separately and the front-side events alone in SC 25 are predicted again. The peak values of front--side RL and halo events have been estimated to be around 31 $\pm$ 3 and 29 $\pm$ 3 respectively. These results are discussed in comparison with the predicted values of sunspots by different authors
\end{abstract}
\keywords{Sun, Solar cycle; Sunspots, Radio-loud CMEs; Halo CMEs}
\end{opening}

\section{Introduction}
     \label{S-Introduction} 
     
	Coronal mass ejections (CMEs) are magnetized plasma materials ejected through the solar corona from the Sun. The Sun is being continuously monitored for CME observations by various space-based coronagraphs. Especially, the Large Angle Spectrometric Coronagraph (LASCO; \citealp{Brueckner1995}) onboard the Solar and Heliospheric Observatory (SOHO; \citealp{Domingo1995}) and the Sun Earth Connection Coronal and Heliospheric Investigation (SECCHI; \citealp{Howard1985}) onboard the Solar Terrestrial Relation Observatory (STEREO; \citealp{Kaiser2008}) are important instruments for white-light observations of CMEs. The properties of these white light CMEs detected by SOHO/LASCO are determined and cataloged online (\citealp{Yashiro2004}, \citealp{Gopalswamy2009}). STEREO/SECCHI/COR2 CMEs have observed and cataloged online (\citealp{Vourlidas2017}). Halo CMEs appears as a halo surrounding the entire disk of the Sun (\citealp{Howard1982}). Nearly 3\% of the total population belongs to this type of CMEs (\citealp{Gopalswamy2010}; \citealp{Lamy2019}). Halo CMEs are usually more geoeffective than other CMEs (\citealp{Lara2006}). Furthermore, it has been widely shown in the literature that front-sided halo CMEs are the most likely to drive severe geomagnetic disturbances (\citealp{Kim2005}; \citealp{Michalek2006}; \citealp{Yermolaev2006}; \citealp{Gopalswamy2007}; \citealp{Scolini2018}).\\

\indent Another group of CMEs more interestingly studied by the solar scientists is radio--loud (RL) CMEs (\citealp{Gopalswamy2012}), because they are generally associated with type II radio bursts and, therefore, with interplanetary shocks. In general, the type II radio emissions are observed as slow drifting features in the radio dynamic spectra. The solar energetic particles (SEPs) are accelerated at the shocks driven by this type of CMEs (\citealp{Kahler1978}; \citealp{Gopalswamy2008}). \citealp{Gopalswamy2019}. (2019) developed a comprehensive catalog that compiles the properties of solar parent eruptions (CME and solar flares) and their associated type II radio bursts. They have used type II radio bursts which are recorded in the decameter-hectometric (DH) domain as the primary data obtained by Wind/WAVES (\citealp{Bougeret1995}, \citealp{Ogilvie1997}) only before October 2006 and by both Wind/WAVES and STEREO/WAVES (\citealp{Bougeret2008}) afterwards. Halo CMEs can be also associated with radio bursts, and such events can be referred to as RL halo CMEs (\citealp{Magdalenic2014}; \citealp{Makela2018}; \citealp{Palmerio2019}). Hence the study of halo CMEs, as well as RL CMEs, is of high interest about concerning space-weather consequences. The connection between sunspot activities of the Sun and CMEs has already been well established (\citealp{Webb1994}; \citealp{Gopalswamy2003}, \citealp{Gopalswamy2010}; \citealp{Robbrecht2009}; \citealp{Mittal2016}; \citealp{Michalek2019}). Although some differences were noted between sunspot and CME rates, the overall similarity between their rates was also observed by \citealp{Gopalswamy2003}.  For example, the energetic populations like the halo and RL CMEs have been found to originate largely from sunspot regions (\citealp{Gopalswamy2010a}) and many quite--Sun filament eruptions can still result in halo CMEs. \citealp{Mittal2016}. (2016) noted that RL halo CMEs follow the solar cycle (SC) variation and \citealp{Michalek2019} found that regular events (CME events of width $>$ 5$^{o}$) track the sunspot cycle. The source location distribution of a large sample of halo CMEs was found to be concentrated near the central meridian (\citealp{Gopalswamy2010a}) and similar to the sunspots' occurrence at low-to-mid latitudes.\\
\indent CMEs that travel in the interplanetary medium and can cause severe geomagnetic effects are identified and cataloged online (\citealp{Cane2003}; \citealp{Richardson2010}; \citealp{Nieves2018}; \citealp{Nieves2019}). Recently, the interplanetary CME (ICME) rate for SC 25 has been predicted by \citealp{Mostl2020} using the correlation of sunspots with ICMEs in the previous two solar cycles. Their study has motivated us to extend it to the prediction of the occurrence rate of halo and RL CMEs for the SC 25. As these are a special group of CMEs, they have been already identified and cataloged online by Coordinated Data Analysis Workshop (CDAW, \url{https://cdaw.gsfc.nasa.gov/}) data center for the previous two solar cycles.\\
\indent Many authors attempted to predict the SC 25 by various methods. For example, \citealp{Gopalswamy2018} studied microwave imaging observations from the Nobeyama Radioheliograph to study the long-term variability such as eruptive prominences and coronal holes from the Sun at low and high latitudes. They found that the microwave brightness temperature in the polar Regions is highly correlated with the polar photospheric magnetic field strength and also it is significantly correlated to the speed of the fast solar wind. They concluded that SC 25 will not be too different from SC 24. \citealp{Bhowmik2018} utilized the magnetic field evolution models for both surface and interior of the Sun. Their predictions indicated that SC 25 is to be similar or slightly stronger than the SC 24 and maximum sunspots will occur around 2024. A non-linear prediction algorithm was used to predict the peak of SC 25 based on time lag and phase space reconstruction by \citealp{Sarp2018} for last five SC. They found that peak of the SC 25 will be at the year 2023.2 $\pm$1.1 with a maximum sunspot number of 154 $\pm$ 12 and also they emphasized that SC 25 is to be stronger than the SC 24.  Similarly \citealp{Bisoi2020} concluded that SC 25 will be relatively stronger than SC 24 and a little weaker than SC 23. But, \citealp{Goncalves2020} showed that SC 25 will be weaker than SC 24 by the Gaussian process technique. They also indicated a continuing trend of declining solar activity as observed in the past two cycles. In contrast, \citealp{McIntosh2020} predicted that SC 25 could be among the strongest SC ever observed. They also added that SC 25 certainly be stronger than present SC 24 and most likely stronger than the previous SC 23. Recently, \citealp{Petrovay2010} and \citealp{Nandy2020} carried out a detailed review for progress in the predictions of SC. They analyzed the results of various authors from different groups based on diverse techniques and performed a comparative analysis of these predictions.\\
\indent In the present paper, we examined the correlations of yearly mean sunspot numbers in the SC 23 and SC 24 with the yearly occurrence rates of halo and RL CMEs. Linear relations between them, utilized to predict the occurrence rates of halo and RL CMEs in SC 25, are also obtained. The data used for these calculations are described in Section 2. Results and discussion are presented in Sections 3 and 4 respectively. A summary is given in Section 5.\\

\section{Data}

The yearly mean sunspot data for the SC 23 (August 1996 -- December 2008) and SC 24 (December 2008 -- December 2019) is obtained from 1997 to 2017 from the online catalog of World Data Center -- Sunspot Index and Long term Solar Observations (WDC--SILSO), Royal Observatory of Belgium, Brussels (\url{http://www.sidc.be/silso/datafiles#total}). The number of halo CMEs and RL CMEs during the above periods are obtained from the SOHO/LASCO halo CME catalog (\url{https://cdaw.gsfc.nasa.gov/CME_list/halo/halo.html}; \citealp{Gopalswamy2009}) and the Wind/WAVES type II bursts and CMEs catalog (\url{https://cdaw.gsfc.nasa.gov/CME_list/radio/waves_type2.html}; \citealp{Gopalswamy2019}), respectively. In these catalogs, the white-light halo CMEs were observed by SOHO/LASCO, and the type II radio bursts were observed by WAVES onboard Wind and STEREO. The type II radio bursts data is available up to September 2017. Hence, we have obtained the number of RL CMEs for the remaining 3 months, together with the halo CMEs for periods corresponding to SOHO/LASCO data gaps, using a simple multiplication factor method described below. Data gaps of duration 3 hours or more in LASCO observations are available online and noted a gap in the year 1998 and 1999.\\

\indent The SOHO/LASCO CME data gaps occurred twice (1998 and 1999) during the SC 23. The SOHO data gap was in 1998 when contact with the whole spacecraft was lost for 3 months 21 days (113 days) which lies from 1998/06/24 11:55 UT to 1998/10/15 19:51 UT. But for the second time, there were around 45 days (1998/12/20 20:30 -- 1999/02/04 21:55). Out of these 45 days, 11 of them belong to the year 1998, and 34 days (1 month 4 days) were in the year 1999. Hence, the total number of days where LASCO was in the data gap is 124 days in the year 1998 and 34 days in the year 1999. We estimated the total number of CMEs (both all RL and halo CMEs, and front-sided RL and halo CMEs) using a multiplying factor that takes the length of the gaps into account. For example in the year 1998, the LASCO catalogue contains 29 halo and 23 RL CMEs that were observed during these 241 days. We derived the multiplying factor by dividing the number of events observed during these days by the number of days observed. For this case, we found that 0.120 (29/241) and 0.095 (23/241) as the multiplying factor for halo and RL CMEs, respectively. We have predicted that the numbers of halo and RL CMEs for this data gap of 124 days are 15 and 12, respectively. So, the total numbers of halo and RL CMEs were increased to 44 and 35 for the year 1998, respectively. In the same method we have also followed for the year 1999. For 331 days of observation, we have obtained the multiplying factors for halo (27) and RL (19) CMEs as 0.082 (27/331) and 0.057 (19/331), respectively. Using these factors, we predicted that the number of halo and RL CMEs for this particular data gap of 34 days as 2 and 2, respectively. So, the total numbers of halo and RL CMEs were raised to 29 and 21, respectively. Please note that the RL CME catalogue (\citealp{Gopalswamy2019}) was available only up to 30/09/2017 in the SC 24. As per the record, 8 RL CMEs were reported for 260 days and the estimated multiplying factor is 0.031 (8/260). With an excess of 3 RL CMEs for the remaining 105 days, the total number of RL CMEs is considered to be 11. The same procedure has been also used to estimate the numbers of front-sided events. In this way, we made all data sets to cover the period from 1997 to 2017 without any data gaps.
 
\section{Results} 
\subsection{Variation of RL and Halo CMEs in SC 23 and SC 24} 

The numbers of RL CMEs and halo CMEs in each year were counted from the data in SC 23 and SC 24 and these are plotted along with the yearly mean sunspot number in Figure 1. The halo CME list of 716 events and the RL CME list of 519 events are tabulated during this study period 1997-2017 in the online catalogs. Further, we have also separated front-side halo and front-side RL CMEs from these sets of halo and RL CMEs, respectively, because some events occurred on the backside of the Sun. Front-side CMEs are important for the space weather conditions on Earth.  The front-side events are considered if the events? source locations are reported within the solar longitude (L) E90$^{o}$ $\le$ L $\le$ W90$^{o}$. Hence, the front-side events were separated from this list and they were counted again and plotted in this Figure 1. Please note that we also estimated the front-side RL and halo CMEs for the data gaps in the years 1998 and 1999 based on the multiplying factor method as already discussed in Section 2. We estimated 0.083 (20/241) and 0.079 (19/241) as the multiplying factors for front side halo and RL CMEs, respectively for the year 1998. We have calculated that the number of front side halo and RL CMEs for this data gap of 124 days is to be 10 for both sets of events. So, the estimated total numbers of front side halo and RL CMEs rose to 30 and 29 for the year 1998, respectively. Similarly in the year 1999, we have also obtained the multiplying factor for front side halo (15) and RL (13) CMEs as 0.045 (15/331) and 0.039 (13/331), respectively. The predicted numbers of front side halo and RL CMEs for 34 days are 2 and 1, respectively. Hence, the numbers of front side halo and RL CMEs were increased to 17 and 14, respectively for the year 1999. Furthermore, there are 7 front side RL CMEs listed for 260 days in the year 2017. The estimated multiplying factor is around 0.027 (7/260) and that gave 3 RL CMEs for the remaining 105 days. So, we have considered a total of 10 front side RL CMEs for the year 2017. There was a total of 424 front-side RL CMEs and 410 front-side halo CMEs. The maximum numbers of RL CMEs are 56 (2001) and 45 (2014) in SC 23 and SC 24, respectively. Similarly, the maximum numbers of front-side RL CMEs are respectively 47 and 28 in SC 23 and SC 24. The ratio of the relative occurrence rate of RL CMEs for SC 24 and SC 23 is 0.52 (182/353). It implies that in SC 24 the occurrence rate of the RL CMEs in 0.52 times lower than that of the SC 23.\\

 \begin{figure} [ht]   
   \centerline{\includegraphics[width=1.\textwidth,clip=]{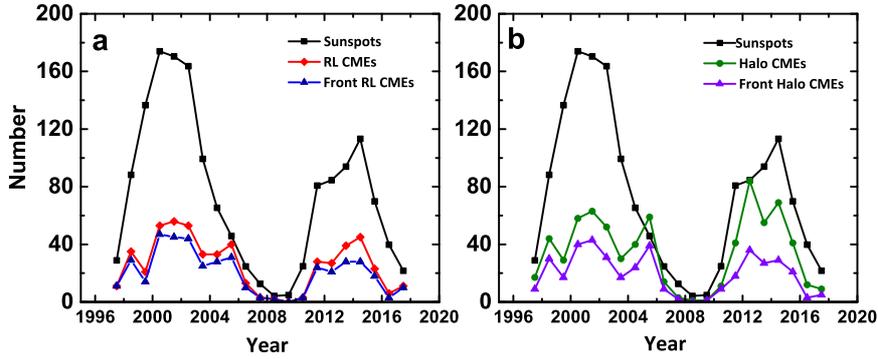}
              }
              \caption{(a) Variation of sunspot number, radio-loud (RL) CMEs and front-side RL CMEs in SC 23 and SC 24. (b) Variation of sunspot number, halo CMEs and front-side halo CMEs in SC 23 and SC 24. SOHO/LASCO data gap was extrapolated}
   \end{figure}
   
On the other hand, the numbers of maximum halo CMEs are 63 (2001) and 84 (2012) in the two cycles. These numbers changed to 43 and 36 when the front-side halo CMEs are only considered in SC 23 and SC 24, respectively. The ratio of the relative occurrence rate of halo CMEs for SC 24 and SC 23 is 0.79 (323/410). The occurrence rate of the RL CMEs in SC 24 is 0.79 times lower than that of the SC 23. From this figure, the occurrence rates of the RL CMEs and halo CMEs follow the mean sunspots number (SSN). Note that the ratios of the relative occurrence rate of RL and halo CMEs did not cover the two final years of SC 24 (2018 and 2019), and so these ratios are not fully accurate.\\

\subsection{Correlation of RL and Halo CMEs with mean sunspots number (SSN)} 

Then the linear relations between the yearly SSN and occurrence rate of halo CMEs and occurrence rate of RL CMEs were obtained separately, and plotted in Figure 2 (a and b). While the correlation between SSN and RL CMEs is good, the one between SSN and halo CMEs is only moderate. As seen in these plots, there are two outliers in each plot that might be due to the occurrence of handful of CMEs from non-sunspot regions (\citealp{Gopalswamy2010}). The correlations improved as shown in Figure 2 (c and d) when the four outliers are removed 

\begin{figure} [ht]   
   \centerline{\includegraphics[width=1.\textwidth,clip=]{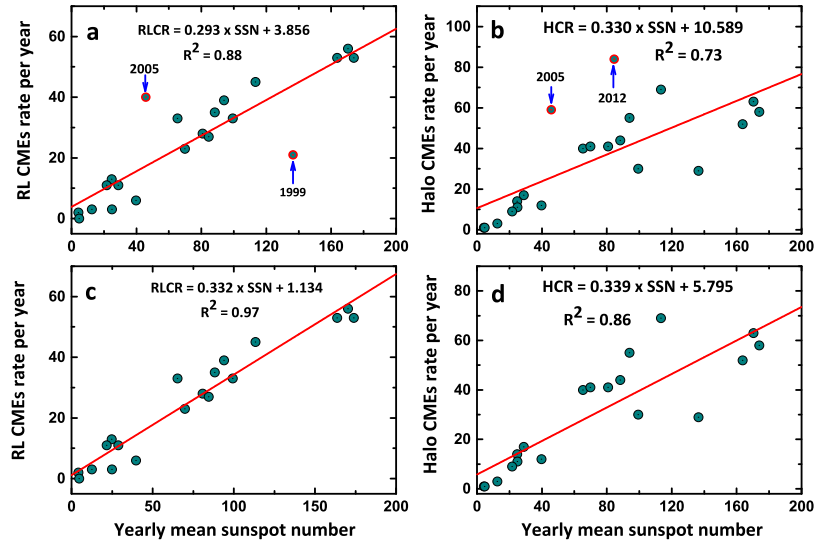}
              }
              \caption{(a) and (b) Correlation of number of sunspots with number of RL CMEs and halo CMEs, respectively. The excluded outliers are marked red circled data points and their corresponding years also presented.  (c) and (d). After the exclusion of the outliers the linear fit is shown by straight red color line. 
                      }
   \end{figure}

\begin{figure} [ht]   
   \centerline{\includegraphics[width=1.\textwidth,clip=]{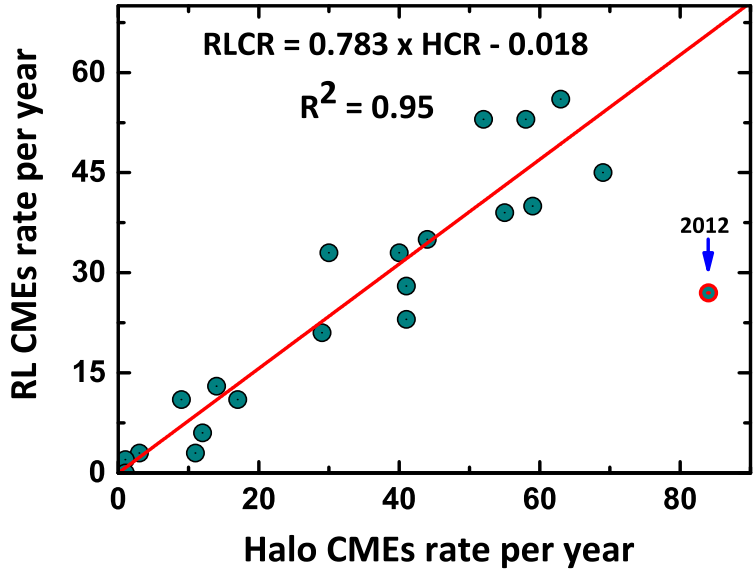}
              }
              \caption{Correlation between yearly mean halo CMEs and yearly mean RL CMEs. The excluded outlier is marked red in color and corresponding year also presented}
   
   \end{figure}

We also found a good correlation between the occurrence rate of halo CMEs and RL CMEs as shown in Figure 3. The Pearson's correlation coefficient is increased from 0.85 to 0.95 when we excluded an outlier. The significance of the correlation is also tested using t-statistic. The p-value is found to be far less than 0.01. It implies that halo CMEs are likely to be faster events that are hence capable of producing type II radio emissions (e.g., \citealp{Reiner2007}).

\subsection{Prediction of RL and Halo CMEs rates}
Following the approach of \citealp{Mostl2020} who predicted the ICME rate for SC 25 based on the SSN, we take the relation between SSN and RL/halo CMEs as a proxy to determine the occurrence rates of RL and halo CMEs during SC 25. In order to obtain the number of sunspots in cycle 25, we considered the prediction values of sunspot numbers by an expert panel of NOAA and NASA in JSON format. The data were collected from the online catalog \url{https://www.swpc.noaa.gov/products/solar-cycle-progression}. The linear relation between yearly SSN and RL CMEs rate (RLCR) is determined as,
\begin{equation}
RLCR = 0.332 \times SSN + 1.134
\end{equation}

and the relation between yearly SSN and halo CMEs occurrence rate  (HCR) is determined as,
\begin{equation}
HCR = 0.339 \times SSN + 5.795
\end{equation}

Using these equations (1) and (2) obtained from the linear fit shown in Figure 2 (c) and (d), respectively as proxies, the number of RL and halo CMEs in SC 25 are estimated and plotted in \label{Figure 4} along with the predicted sunspot number.\\
According to the predicted values of the sunspot number of NOAA, the solar maximum in SC 25 is expected to be in July 2025 with a peak of 115. As given in the prediction data of 1$\sigma$ uncertainty, the peak may go up to a maximum of 125 and may go low up to 105. Using these sunspot data in equations 1 and 2, the numbers of RL and halo CMEs are predicated for SC 25. Correspondingly, the 25th solar maximum would likely to produce 39 RL (45 halos) CMEs. Including the 1$\sigma$  uncertainty values of sunspot data, these numbers may vary from 43 RL (49 halos) to 36 RL (41 halos).

\begin{figure} [ht]   
   \centerline{\includegraphics[width=1.\textwidth,clip=]{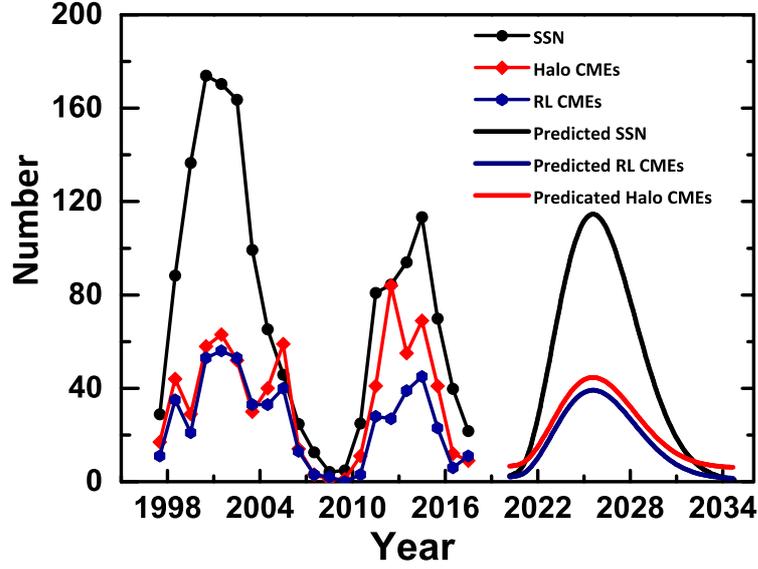}
              }
              \caption{Predicted numbers of sunspot, RL-CMEs, halo CMEs in SC 25 are drawn along with the SC 23 and SC 24}
 
   \end{figure}
   
Further, the backside events in the data of RL CMEs and halo CMEs were removed to get a clear linear relationship between the front-side RL CMEs and front-side halo CMEs with SSN without any uncertainty.  Before exclusion of these outliers, we found that Pearson?s correlation coefficient are 0.89 and 0.79 for front-side RL and halo CMEs, respectively (linear fitted lines are shown in blue in color).  Now the correlations were obtained again as shown in Figure 5. The coefficient of correlation remains as high as 0.9 after excluding the outliers from RL CMEs and Halo CMEs.  The linear relation between yearly SSN and front-side RL CMEs occurrence rate (FRLCR) is determined as,
\begin{equation}
FRLCR = 0.266 \times SSN + 0.844
\end{equation}

and in the same way the relation between yearly SSN and front-side halo CMEs occurrence rate (FHCR) is determined as,
\begin{equation}
FHCR = 0.231 \times SSN + 2.038	
\end{equation}

\begin{figure} [ht]   
   \centerline{\includegraphics[width=1.\textwidth,clip=]{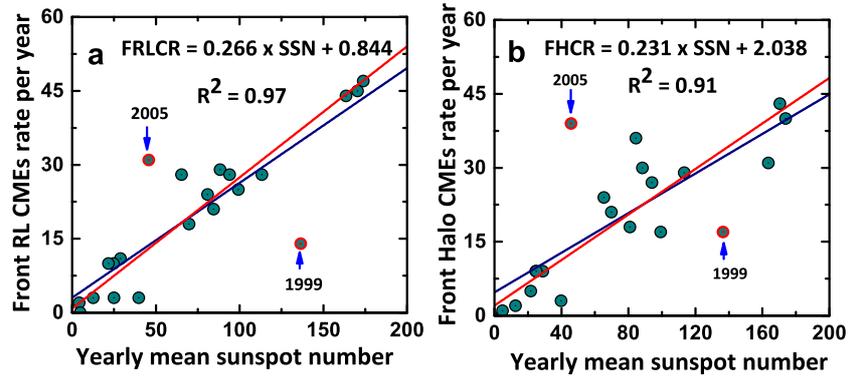}
              }
              \caption{(a) Correlation between number of sunspots and number of front-side RL CMEs (FRLCR), (b) Correlation between SSN and number of front-side halo CMEs (FHCR). Linear fits are shown before and after the exclusion of the outliers by blue and red color straight lines, respectively.}
  
   \end{figure}

\begin{figure} [ht]   
   \centerline{\includegraphics[width=1.\textwidth,clip=]{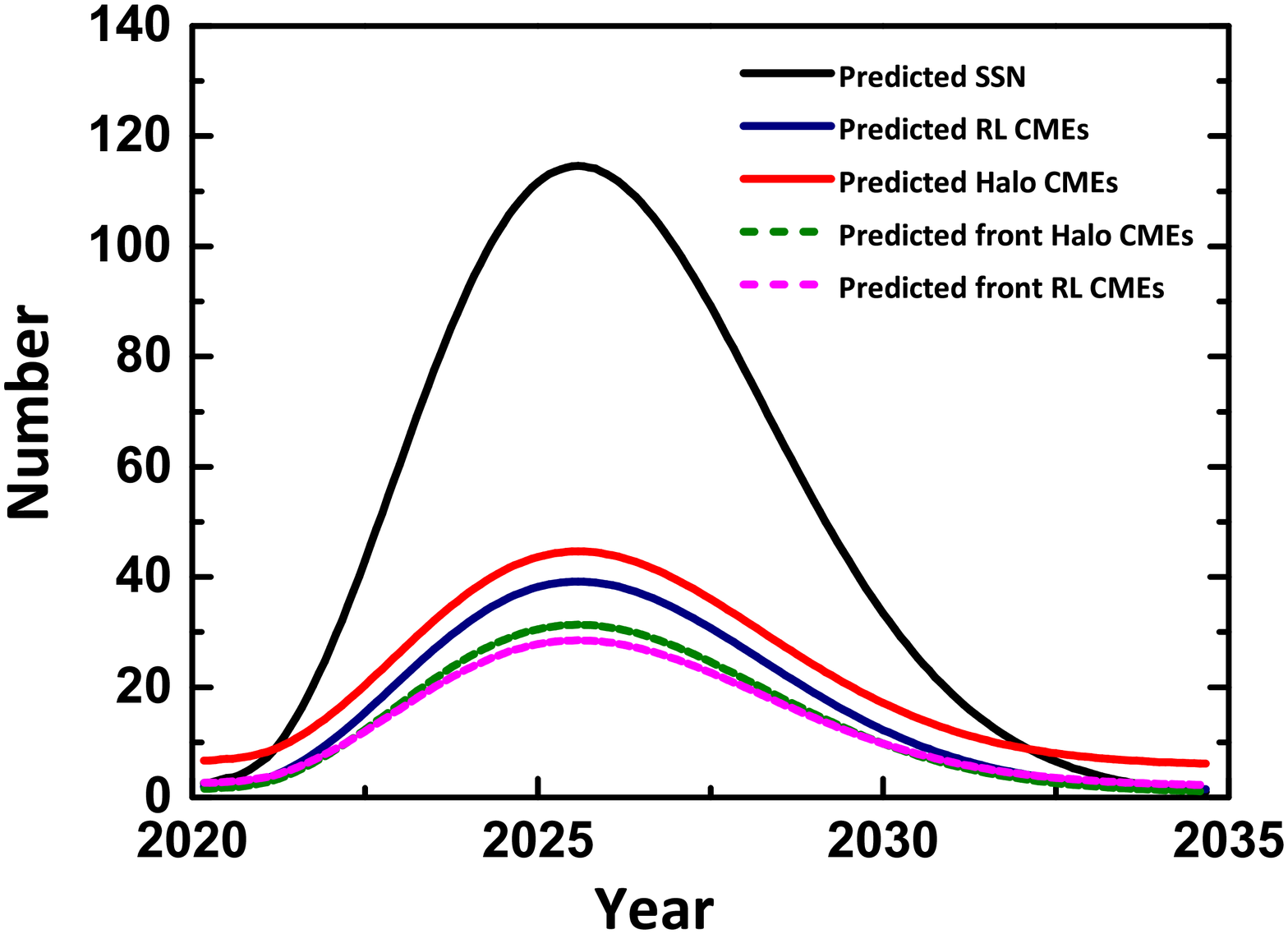}
              }
              \caption{Predicted numbers of sunspots, RL-CMEs, halo CMEs, front-side RL CMEs and front-side halo CMEs in SC 25.}
 
   \end{figure}

Using these relations (3) and (4) as proxy, we determined the number of front-sided RL and halo CMEs in SC 25. They are plotted in Figure 6 along with the predicted RL CMEs, predicted halo CMEs, and predicted SSN.\\
Now according to the new estimates using equations (3) and (4), the numbers of front-side RL and front-side halo CMEs during solar maximum would be around 31 and 29, respectively. Including the 1$\sigma$ uncertainty in sunspot number, these numbers may vary from 34 RL (31 halos) to 29 RL (26 halos) CMEs.\\

\section{Discussion} 

Recently, \citealp{Bhowmik2018} predicted the strength and timing of SC 25 using the magnetic field evolution model. It was estimated that the maximum number of sunspots in solar maximum is 118 in 2024 $\pm$1 and the range would be 139 -- 109. This peak value is nearly similar to that (115, provided by NOAA) used in this paper. Also the range they predicted 139 -- 109 is near to that of NOAA (125 -- 105). We have adopted this SSN range and predicted the occurrence rate of RL CMEs, halo CMEs, front-side RL CMEs and front-side halo CMEs by using our proxy equations 1, 2, 3 and 4, respectively. For the minimum SSN (109), we found the occurrence rate of RL CME, halo CMEs, front-side RL CMEs and front-side halo CMEs are 37, 43, 30 and 27, respectively. We have also estimated for the maximum SSN (139) for RL CMEs, halo CMEs, front-side RL CMEs and front-side halo CMEs. The occurrence rates of these CMEs are 47, 53, 38 and 34, respectively.\\

Prediction of sunspot by \citealp{Goncalves2020} using a different method of Warped Gaussian process regression gave a similar peak of sunspot number 117 in the year 2024. On the other hand, \citealp{Gopalswamy2018} predicted a greater strength of SC 25 using microwave brightness observations with a total sunspot number of 148 during maximum. A relatively stronger peak ($~$131 to 134 $\pm$ 11) than SC 24 has been estimated recently by \citealp{Bisoi2020}.  However, \citealp{Sarp2018} predicted a peak sunspot number of 154 $\pm$12 around the year 2023.2 using a non-linear based algorithm. A higher peak of sunspot number 232, different from the above reports, has been predicted by \citealp{McIntosh2020} using some terminator type events. In spite of all the above deviations, the prediction of halo CMEs and radio-loud CMEs will be of great use to advance the forecasting models of space-weather consequences.\\
\citealp{Gopalswamy2015} reported that the fraction of halo CMEs in SC 23 raised to in SC 24. This was explained by them as an anomalous expansion of CMEs due to heliospheric conditions. So the number of halo CMEs predicted for SC 25 in the present paper may change upon the heliospheric conditions. However, \citealp{Lamy2019} reported that the CME rates were relatively larger during SC 24 than during SC 23 and they also reported a significant variation in the identification of the halo CMEs and their reports in different CME catalogs.  On the other hand, \citealp{Gopalswamy2019} presented that the total number 296 of RL CMEs in SC 23 dropped to 179 in SC 24. Also, they reported more or less similar percentage drop in sunspot number in SC 24 from SC 23. Hence the number of RL CMEs predicted for SC 25 is likely to be similar to that of SC 24.\\
    
\section{Summary} 

In the present paper, the occurrence rates of radio-loud (RL) and halo CMEs in SC 25 are predicted using the correlations of the numbers of these events each year with the solar cycle indicator (yearly mean sunspot number) in the previous two cycles. For this analysis, the annual mean sunspot number and the data of RL CMEs and halo CMEs were obtained for the SC 23 and SC 24 from the respective online catalogs. First, the linear relations between the two sets (number of sunspots and number of halo CMEs; the number of sunspots and number of RL CMEs) are found from the correlation analysis. The correlation coefficient ($>$ 0.9) found in this analysis showed a very good correlation in the two sets.  Using the predicted values of sunspot number by NOAA/NASA for SC 25 and utilizing the above relations, the occurrence rates of RL and halo CMEs are determined for the entire SC 25. As per the records of NOAA, the solar maximum in cycle 25 is expected to be in July 2025 with a peak of 115 sunspots. For this peak sunspot number during the solar maximum, we obtained a total of 39 RL and 45 halo CMEs. After neglecting the backside events of RL and halo CMEs, the relations are obtained again using only the front-side RL and halo events. Correspondingly, the numbers of front-side RL and halo events are found to be 31 and 29, respectively. The sunspot numbers in solar maximum of cycle 25 reported by several authors were also considered and discussed in comparison with the earlier results. Many authors have pointed out a similar or slightly higher strength for cycle 25 to that of cycle 24 except \citealp{McIntosh2020}.

\begin{acks}
 We thank the reviewer for useful constructive comments to improve the quality of this manuscript. We greatly acknowledge the data support provided by various online data centers of NOAA and NASA. The SOHO/LASCO CME catalog is generated and maintained at the CDAW Data Center by NASA and The Catholic University of America in cooperation with the Naval Research Laboratory. SOHO is a project of international cooperation between ESA and NASA. We also thank Sunspot data from the World Data Center SILSO, Royal Observatory of Belgium, Brussels for their open data policies. O. Prakash thanks to the Chinese Academy of Sciences for providing General Financial Grant from the China Postdoctoral Science Foundation. 
\end{acks}

 \section*{Conflict of interest}
The authors declare that they have no conflict of interest.

\end{article} 


\begin{thebibliography}{}

\bibitem[\protect\citeauthoryear{{Bhowmik} and {Nandy}}{2018}]{Bhowmik2018}Bhowmik, P. and Nandy, D.: 2018, {\it Nature Communications} {\bf 9}, 5209. doi:10.1038/s41467-018-07690-0.

\bibitem[\protect\citeauthoryear{{Bisoi}, {Janardhan}, and {Ananthakrishnan}}{2020}]{Bisoi2020}Bisoi, S.K., Janardhan, P., and Ananthakrishnan, S.: 2020, {\it Journal of Geophysical Research (Space Physics)} {\bf 125}, e27508. doi:10.1029/2019JA027508.

\bibitem[\protect\citeauthoryear{{Bougeret \emph{et al.}}}{1995}]{Bougeret1995}Bougeret, J.-L., Kaiser, M.L., Kellogg, P.J., Manning, R., Goetz, K., Monson, S.J., and, ...: 1995, {\it Space Science Reviews} {\bf 71}, 231. doi:10.1007/BF00751331.

\bibitem[\protect\citeauthoryear{{Bougeret} \emph{et al.}}{2008}]{Bougeret2008}Bougeret, J.L., Goetz, K., Kaiser, M.L., Bale, S.D., Kellogg, P.J., Maksimovic, M., and, ...: 2008, {\it Space Science Reviews} {\bf 136}, 487. doi:10.1007/s11214-007-9298-8.

\bibitem[\protect\citeauthoryear{{Brueckner \emph{et al.}}}{1995}]{Brueckner1995} Brueckner, G.E., Howard, R.A., Koomen, M.J., Korendyke, C.M., Michels, D.J., Moses, J.D., and, ...: 1995, {\it Solar Physics} {\bf 162}, 357. doi:10.1007/BF00733434.

\bibitem[\protect\citeauthoryear{{Cane} and {Richardson}}{2003}]{Cane2003}Cane, H.V. and Richardson, I.G.: 2003, {\it Journal of Geophysical Research (Space Physics)} {\bf 108}, 1156. doi:10.1029/2002JA009817.

\bibitem[\protect\citeauthoryear{{Domingo}, {Fleck}, and {Poland}}{1995}]{Domingo1995}Domingo, V., Fleck, B., and Poland, A.I.: 1995, {\it Solar Physics} {\bf 162}, 1. doi:10.1007/BF00733425.

\bibitem[\protect\citeauthoryear{{Gon{\c{c}}alves}, {Echer}, and {Frigo}}{2020}]{Goncalves2020}Gon{\c{c}}alves, {\'I}.G., Echer, E., and Frigo, E.: 2020, {\it Advances in Space Research} {\bf 65}, 677. doi:10.1016/j.asr.2019.11.011.

\bibitem[\protect\citeauthoryear{{Gopalswamy} \emph{et al.}}{2003}]{Gopalswamy2003}Gopalswamy, N., Lara, A., Yashiro, S., and Howard, R.A.: 2003, {\it The Astrophysical Journal} {\bf 598}, L63. doi:10.1086/380430.

\bibitem[\protect\citeauthoryear{{Gopalswamy}, {Yashiro}, and {Akiyama}}{2007}]{Gopalswamy2007}Gopalswamy, N., Yashiro, S., and Akiyama, S.: 2007, {\it Journal of Geophysical Research (Space Physics)} {\bf 112}, A06112. doi:10.1029/2006JA012149.

\bibitem[\protect\citeauthoryear{{Gopalswamy \emph{et al.}}}{2008}]{Gopalswamy2008}Gopalswamy, N., Yashiro, S., Akiyama, S., M{\"a}kel{\"a}, P., Xie, H., Kaiser, M.L., and, ...: 2008, {\it Annales Geophysicae} {\bf 26}, 3033. doi:10.5194/angeo-26-3033-2008.

\bibitem[\protect\citeauthoryear{{Gopalswamy} \emph{et al.}}{2009}]{Gopalswamy2009}Gopalswamy, N., Yashiro, S., Michalek, G., Stenborg, G., Vourlidas, A., Freeland, S., and, ...: 2009, {\it Earth Moon and Planets} {\bf 104}, 295. doi:10.1007/s11038-008-9282-7.

\bibitem[\protect\citeauthoryear{{Gopalswamy} and {Davila}}{2010}]{Gopalswamy2010}Gopalswamy, N. and Davila, J.M.: 2010, {\it 20th National Solar Physics Meeting} {\bf 20}, 160.

\bibitem[\protect\citeauthoryear{{Gopalswamy} \emph{et al.}}{2010}]{Gopalswamy2010a}Gopalswamy, N., Akiyama, S., Yashiro, S., and M{\"a}kel{\"a}, P.: 2010, {\it Astrophysics and Space Science Proceedings} {\bf 19}, 289. doi:10.1007/978-3-642-02859-5\_24.

\bibitem[\protect\citeauthoryear{{Gopalswamy} \emph{et al.}}{2012}]{Gopalswamy2012}Gopalswamy, N., M{\"a}Kel{\"a}, P., Akiyama, S., Yashiro, S., Xie, H., MacDowall, R.J., and, ...: 2012, {\it Journal of Geophysical Research (Space Physics)} {\bf 117}, A08106. doi:10.1029/2012JA017610.

\bibitem[\protect\citeauthoryear{{Gopalswamy}, {Tsurutani}, and {Yan}}{2015}]{Gopalswamy2015}Gopalswamy, N., Tsurutani, B., and Yan, Y.: 2015, {\it Progress in Earth and Planetary Science} {\bf 2}, 13. doi:10.1186/s40645-015-0043-8.

\bibitem[\protect\citeauthoryear{{Gopalswamy} \emph{et al.}}{2018}]{Gopalswamy2018}Gopalswamy, N., M{\"a}kel{\"a}, P., Yashiro, S., and Akiyama, S.: 2018, {\it Journal of Atmospheric and Solar-Terrestrial Physics} {\bf 176}, 26. doi:10.1016/j.jastp.2018.04.005.

\bibitem[\protect\citeauthoryear{{Gopalswamy}, {M{\"a}kel{\"a}}, and {Yashiro}}{2019}]{Gopalswamy2019}Gopalswamy, N., M{\"a}kel{\"a}, P., and Yashiro, S.: 2019, {\it Sun and Geosphere} {\bf 14}, 111. doi:10.31401/SunGeo.2019.02.03.

\bibitem[\protect\citeauthoryear{{Howard} \emph{et al.}}{1982}]{Howard1982}Howard, R.A., Michels, D.J., Sheeley, N.R., and Koomen, M.J.: 1982, {\it The Astrophysical Journal} {\bf 263}, L101. doi:10.1086/183932.

\bibitem[\protect\citeauthoryear{{Howard} \emph{et al.}}{1985}]{Howard1985}Howard, R.A., Sheeley, N.R., Michels, D.J., and Koomen, M.J.: 1985, {\it Journal of Geophysical Research} {\bf 90}, 8173. doi:10.1029/JA090iA09p08173.

\bibitem[\protect\citeauthoryear{{Kahler}, {Hildner}, and {Van Hollebeke}}{1978}]{Kahler1978}Kahler, S.W., Hildner, E., and Van Hollebeke, M.A.I.: 1978, {\it Solar Physics} {\bf 57}, 429. doi:10.1007/BF00160116.

\bibitem[\protect\citeauthoryear{{Kaiser} \emph{et al.}}{2008}]{Kaiser2008}Kaiser, M.L., Kucera, T.A., Davila, J.M., St. Cyr, O.C., Guhathakurta, M., and Christian, E.: 2008, {\it Space Science Reviews} {\bf 136}, 5. doi:10.1007/s11214-007-9277-0.

\bibitem[\protect\citeauthoryear{{Kim} \emph{et al.}}{2005}]{Kim2005}Kim, R.-S., Cho, K.-S., Moon, Y.-J., Kim, Y.-H., Yi, Y., Dryer, M., and, ...: 2005, {\it Journal of Geophysical Research (Space Physics)} {\bf 110}, A11104. doi:10.1029/2005JA011218.

\bibitem[\protect\citeauthoryear{{Lamy} \emph{et al.}}{2019}]{Lamy2019}Lamy, P.L., Floyd, O., Boclet, B., Wojak, J., Gilardy, H., and Barlyaeva, T.: 2019, {\it Space Science Reviews} {\bf 215}, 39. doi:10.1007/s11214-019-0605-y.

\bibitem[\protect\citeauthoryear{{Lara} \emph{et al.}}{2006}]{Lara2006}Lara, A., Gopalswamy, N., Xie, H., Mendoza-Torres, E., P{\'e}Rez-Er{\'\i}Quez, R., and Michalek, G.: 2006, {\it Journal of Geophysical Research (Space Physics)} {\bf 111}, A06107. doi:10.1029/2005JA011431.

\bibitem[\protect\citeauthoryear{{Magdaleni{\'c}} \emph{et al.}}{2014}]{Magdalenic2014}Magdaleni{\'c}, J., Marqu{\'e}, C., Krupar, V., Mierla, M., Zhukov, A.N., Rodriguez, L., and, ...: 2014, {\it The Astrophysical Journal} {\bf 791}, 115. doi:10.1088/0004-637X/791/2/115.

\bibitem[\protect\citeauthoryear{{M{\"a}kel{\"a}}, {Gopalswamy}, and {Akiyama}}{2018}]{Makela2018}M{\"a}kel{\"a}, P., Gopalswamy, N., and Akiyama, S.: 2018, {\it The Astrophysical Journal} {\bf 867}, 40. doi:10.3847/1538-4357/aae2b6.

\bibitem[\protect\citeauthoryear{{McIntosh} \emph{et al.}}{2020}]{McIntosh2020}McIntosh, S.W., Chapman, S., Leamon, R.J., Egeland, R., and Watkins, N.W.: 2020, {\it Solar Physics} {\bf 295}, 163. doi:10.1007/s11207-020-01723-y.

\bibitem[\protect\citeauthoryear{{Michalek} \emph{et al.}}{2006}]{Michalek2006}Michalek, G., Gopalswamy, N., Lara, A., and Yashiro, S.: 2006, {\it Space Weather} {\bf 4}, S10003. doi:10.1029/2005SW000218.

\bibitem[\protect\citeauthoryear{{Michalek}, {Gopalswamy}, and {Yashiro}}{2019}]{Michalek2019}Michalek, G., Gopalswamy, N., and Yashiro, S.: 2019, {\it The Astrophysical Journal} {\bf 880}, 51. doi:10.3847/1538-4357/ab26a7.

\bibitem[\protect\citeauthoryear{{Mittal} \emph{et al.}}{2016}]{Mittal2016}Mittal, N., Sharma, J., Verma, V.K., and Garg, V.: 2016, {\it New Astronomy} {\bf 47}, 64. doi:10.1016/j.newast.2016.02.004.

\bibitem[\protect\citeauthoryear{{M{\"o}stl} \emph{et al.}}{2020}]{Mostl2020}M{\"o}stl, C., Weiss, A.J., Bailey, R.L., Reiss, M.A., Amerstorfer, T., Hinterreiter, J., and, ...: 2020, {\it The Astrophysical Journal} {\bf 903}, 92. doi:10.3847/1538-4357/abb9a1.

\bibitem[\protect\citeauthoryear{{Nandy}}{2020}]{Nandy2020}Nandy, D.: 2020, {\it arXiv e-prints}, arXiv:2009.01908.

\bibitem[\protect\citeauthoryear{{Nieves-Chinchilla} \emph{et al.}}{2018}]{Nieves2018}Nieves-Chinchilla, T., Vourlidas, A., Raymond, J.C., Linton, M.G., Al-haddad, N., Savani, N.P., and, ...: 2018, {\it Solar Physics} {\bf 293}, 25. doi:10.1007/s11207-018-1247-z.

\bibitem[\protect\citeauthoryear{{Nieves-Chinchilla} \emph{et al.}}{2019}]{Nieves2019}Nieves-Chinchilla, T., Jian, L.K., Balmaceda, L., Vourlidas, A., dos Santos, L.F.G., and Szabo, A.: 2019, {\it Solar Physics} {\bf 294}, 89. doi:10.1007/s11207-019-1477-8.

\bibitem[\protect\citeauthoryear{{Ogilvie} and {Desch}}{1997}]{Ogilvie1997}Ogilvie, K.W. and Desch, M.D.: 1997, {\it Advances in Space Research} {\bf 20}, 559. doi:10.1016/S0273-1177(97)00439-0.

\bibitem[\protect\citeauthoryear{{Palmerio} \emph{et al.}}{2019}]{Palmerio2019}Palmerio, E., Scolini, C., Barnes, D., Magdaleni{\'c}, J., West, M.J., Zhukov, A.N., and, ...: 2019, {\it The Astrophysical Journal} {\bf 878}, 37. doi:10.3847/1538-4357/ab1850.

\bibitem[\protect\citeauthoryear{{Petrovay}}{2010}]{Petrovay2010}Petrovay, K.: 2010, {\it Living Reviews in Solar Physics} {\bf 7}, 6. doi:10.12942/lrsp-2010-6.

\bibitem[\protect\citeauthoryear{{Reiner}, {Kaiser}, and {Bougeret}}{2007}]{Reiner2007}Reiner, M.J., Kaiser, M.L., and Bougeret, J.-L.: 2007, {\it The Astrophysical Journal} {\bf 663}, 1369. doi:10.1086/518683.

\bibitem[\protect\citeauthoryear{{Richardson} and {Cane}}{2010}]{Richardson2010}Richardson, I.G. and Cane, H.V.: 2010, {\it AGU Fall Meeting Abstracts}.

\bibitem[\protect\citeauthoryear{{Robbrecht}, {Berghmans}, and {Van der Linden}}{2009}]{Robbrecht2009}Robbrecht, E., Berghmans, D., and Van der Linden, R.A.M.: 2009, {\it The Astrophysical Journal} {\bf 691}, 1222. doi:10.1088/0004-637X/691/2/1222.

\bibitem[\protect\citeauthoryear{{Sarp} \emph{et al.}}{2018}]{Sarp2018}Sarp, V., Kilcik, A., Yurchyshyn, V., Rozelot, J.P., and Ozguc, A.: 2018, {\it Monthly Notices of the Royal Astronomical Society} {\bf 481}, 2981. doi:10.1093/mnras/sty2470.

\bibitem[\protect\citeauthoryear{{Scolini} \emph{et al.}}{2018}]{Scolini2018}Scolini, C., Messerotti, M., Poedts, S., and Rodriguez, L.: 2018, {\it Journal of Space Weather and Space Climate} {\bf 8}, A9. doi:10.1051/swsc/2017046.

\bibitem[\protect\citeauthoryear{{Vourlidas} \emph{et al.}}{2017}]{Vourlidas2017}Vourlidas, A., Balmaceda, L.A., Stenborg, G., and Dal Lago, A.: 2017, {\it The Astrophysical Journal} {\bf 838}, 141. doi:10.3847/1538-4357/aa67f0.

\bibitem[\protect\citeauthoryear{{Webb} and {Howard}}{1994}]{Webb1994}Webb, D.F. and Howard, R.A.: 1994, {\it Journal of Geophysical Research} {\bf 99}, 4201. doi:10.1029/93JA02742.

\bibitem[\protect\citeauthoryear{{Yashiro} \emph{et al.}}{2004}]{Yashiro2004}Yashiro, S., Gopalswamy, N., Michalek, G., St. Cyr, O.C., Plunkett, S.P., Rich, N.B., and, ...: 2004, {\it Journal of Geophysical Research (Space Physics)} {\bf 109}, A07105. doi:10.1029/2003JA010282.

\bibitem[\protect\citeauthoryear{{Yermolaev} and {Yermolaev}}{2006}]{Yermolaev2006}Yermolaev, Y.I. and Yermolaev, M.Y.: 2006, {\it Advances in Space Research} {\bf 37}, 1175. doi:10.1016/j.asr.2005.03.130.




\end{thebibliography}
\end{document}